\documentclass[aps,prl,twocolumn,superscriptaddress,showpacs]{revtex4}
\usepackage{amsmath}
\usepackage{amssymb}
\usepackage{graphicx}
\usepackage{dcolumn}
\usepackage{bm}

\begin{document}

\title{Observation of a Dirac point in microwave experiments with a photonic crystal modeling graphene}

\author{S.~Bittner}
\affiliation{Institut f{\"u}r Kernphysik, Technische Universit{\"a}t
Darmstadt, D-64289 Darmstadt, Germany}

\author{B.~Dietz}
\affiliation{Institut f{\"u}r Kernphysik, Technische Universit{\"a}t
Darmstadt, D-64289 Darmstadt, Germany}

\author{M.~Miski-Oglu}
\affiliation{Institut f{\"u}r Kernphysik, Technische Universit{\"a}t
Darmstadt, D-64289 Darmstadt, Germany}

\author{P.~Oria Iriarte}
\affiliation{Institut f{\"u}r Kernphysik, Technische Universit{\"a}t
Darmstadt, D-64289 Darmstadt, Germany}

\author{A.~Richter}
\email{richter@ikp.tu-darmstadt.de}
\affiliation{Institut f{\"u}r
Kernphysik, Technische Universit{\"a}t Darmstadt, D-64289 Darmstadt,
Germany}
\affiliation{$\rm ECT^*$, Villa Tambosi, I-38123 Villazzano (Trento), Italy}

\author{F.~Sch{\"a}fer}
\affiliation{LENS, University of Florence, I-50019
Sesto-Fiorentino (Firenze), Italy}

\date{\today}

\begin{abstract}
We present measurements of transmission and reflection spectra of a microwave photonic crystal composed of 874 metallic cylinders arranged in a triangular lattice. The spectra show clear evidence of a Dirac point, a characteristic of a spectrum of relativistic massless fermions.  In fact, Dirac points are a peculiar property of the electronic band structure of graphene, whose properties consequently can be described by the relativistic Dirac equation. In the vicinity of the Dirac point, the measured reflection spectra resemble those obtained by conductance measurements in scanning tunneling microscopy of graphene flakes.
\end{abstract}

\maketitle

Graphene is a monolayer of carbon atoms arranged in a honeycomb lattice \cite{Novoselov2004,Novoselov2005}. Due to its peculiar electronic properties this carbon allotrope recently attracted a lot of attention in condensed matter physics. The conduction and the valence band of the electronic energy in graphene form conically shaped valleys that touch each other at the corners of the Brillouin zone \cite{Wallace1947}. As a consequence, close to these touch points the energy of the electron depends linearly on its quasi-momentum vector. This linear dispersion relation implies an energy independent velocity, and the related wave equation is the Dirac equation \cite{Semenoff1984}. Thus, although the Fermi velocity is typically 300 times smaller than that of light, the energy spectrum of the electrons in graphene is similar to that of massless relativistic fermions \cite{Semenoff1984,Katsnelson2007,Beenakker2008,Castro2009}. However, graphene with its Dirac spectrum is not an exception. Photonic crystals, an optical analogue of ordinary crystals, possess similar particular properties \cite{Parimi2004,Sepkhanov2007}. The unusual transmission properties near a Dirac point predicted in \cite{Sepkhanov2007} were observed experimentally both in sonic and in microwave photonic crystals \cite{Zhang2008,Zandbergen2010}.

\begin{figure}[t]
{\includegraphics[width=\linewidth]{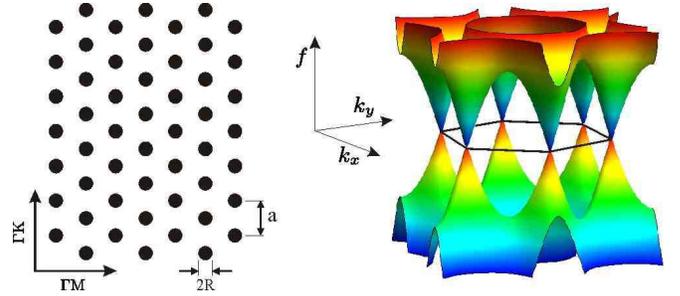}}
\caption{(Color online) Left: Triangular lattice of metallic cylinders. The arrows indicate the two different directions $\Gamma\rm{M}$ and $\Gamma\rm{K}$ in the triangular lattice; the radius $R$ of the metallic cylinders equals $R=0.25a$, where $a$ is the lattice constant. Right: The plot of the numerically determined band structure $f(k_x,k_y)$, as function of the quasi momentum components $(k_x,k_y)$. The solid lines indicate the hexagonal Brillouin zone.
} 
\label{fig1}
\end{figure} 
The photonic crystal considered in the present work is two-dimensional and composed of rows of metallic cylinders which are arranged to form the triangular lattice schematically shown in the left part of Fig.~\ref{fig1}. Electromagnetic waves propagating in such a periodic structure exhibit a dispersion relation with a band structure similar to the electronic band structure in a solid. That for the triangular lattice of metallic cylinders with radius $R=0.25a$, where $a$ is the lattice constant, is shown on the right part of Fig.~\ref{fig1}. It was obtained by solving numerically the Helmholtz equation with the finite difference method \cite{Smirnova2002}. In this so-called band diagram the frequency $f$ is plotted as function of the quasi momentum components $(k_x,k_y)$ for the first two propagating modes. Around the corners of the first Brillouin zone the bands have the shape of cones and the band structure resembles that of the electronic energy in graphene \cite{Novoselov2005,Katsnelson2007}. In fact both systems have the same Bravais lattice and the wave functions with quasi momentum at the corners of the first Brillouin zone possess the same symmetry group \cite{Slonczewski1958}. In the literature the touching point is referred to as Dirac point \cite{Novoselov2005,Raghu2008} and its frequency as the Dirac frequency. 

In the vicinity of a touching point the Hamiltonian governing the spectrum can be written as a $2\times2$ matrix in the basis of the doublet states which are degenerate at the Dirac frequency and has the form \cite{Raghu2008}
\begin{equation}
\hat H=\omega_D\openone+v_D\left({\delta k}_x\hat\sigma_x+{\delta k}_y\hat\sigma_y\right).
\label{HeffDirac}
\end{equation}
Here, $\hat\sigma_x$, $\hat\sigma_y$ are the Pauli matrices, $\delta \vec{k}=({\delta k}_x,{\delta k}_y)$ is the displacement vector with respect to a corner of the first Brillouin zone, $\omega=2\pi f$, with $f$ denoting the excitation frequency, $\omega_D$ is the Dirac frequency and $v_D=\vert\vec\nabla\omega(\vec{k})\vert$ is the group velocity which is approximately constant close to the Dirac point. Diagonalization of $\hat H$ in Eq.~(\ref{HeffDirac}) yields
\begin{equation}
\omega(\vert\delta\vec{k}\vert )=\omega_D\pm v_D\sqrt{{\delta k}_x^2+{\delta k}_y^2},
 \label{DiracDispersion}
\end{equation}
which reproduces the linear dispersion relation around the tips of the cones of the Brillouin zones shown in Fig.~\ref{fig1}. The substitution ${\delta k}_x\rightarrow -i \partial_x$ and ${\delta k}_y\rightarrow -i \partial_y$ in Eq.~(\ref{HeffDirac}) leads to the Dirac equation 
\begin{equation}
\begin{array}{ccccc}
\left( 
\begin{array}{cc}
 0& \partial_x-i \partial_y\\
\partial_x+i \partial_y&0\\
\end{array}
\right) & 
\displaystyle{\psi_1\choose \psi_2}&=& \displaystyle i\, \frac{\omega-\omega_D}{v_D}&\displaystyle{\psi_1\choose \psi_2}\\
\end{array},
\label{DiracEq}
\end{equation}
where  $\psi_1,\, \psi_2$ represent the amplitudes of the degenerate doublet state at the Dirac frequency. Thus, despite of the fact that photons are bosons in the vicinity of $\omega_D$ the waves in a photonic crystal can be effectively described by the Dirac equation for fermions with spin $1/2$.

\begin{figure}[t]
{\includegraphics[width=0.7\linewidth]{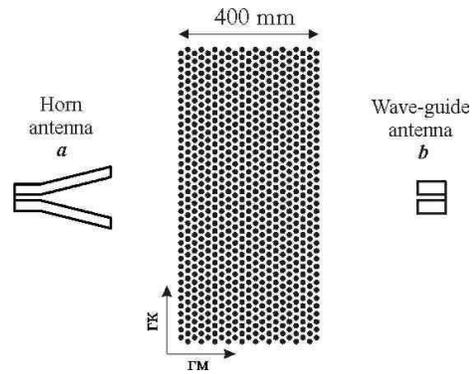}}
\caption{Scheme of the experimental setup. The photonic crystal consists of altogether $874$ metallic cylinders arranged in a triangular lattice. Electromagnetic waves are transmitted from the horn antenna through the lattice to the wave guide antenna. The antennae and the lattice are oriented with respect to each other such that the waves impinge the lattice in $\Gamma\rm{M}$ direction (see Fig.\ref{fig1}). } 
\label{fig3}
\end{figure}

The similarity of the band structure of a macroscopic photonic crystal with the electronic band structure of graphene, which is experimentally much more difficult to access, allows the experimental study of various relativistic phenomena. The photonic crystal used in the experiments is composed of a total of 874 metallic cylinders with radius $R=5$~mm and height $h=8$~mm squeezed between two metallic plates. The experimental setup is shown in Fig.~\ref{fig3}. Each cylinder is screwed to the top and bottom copper plate to ensure a proper electric contact even at high frequencies and thus reproducibility of the measurements. The resulting photonic crystal consists of $38\times23$ cylinders and has a size of $400\times900$~mm. 
Below a certain excitation frequency $f_c=c/2h$, with $c$ being the speed of light and $h$ the gap between the plates, only the lowest transverse magnetic (TM) mode with the electric field perpendicular to the plates can propagate. Consequently, like for two-dimensional microwave billiards \cite{Richter1999,StoeckmannBuch2000}, the propagating modes correspond to solutions of the two-dimensional scalar Helmholtz equation, which is mathematically identical to the Schr\"odinger equation for the corresponding quantum multiple-scattering problem.

To determine the band structure of the photonic crystal, transmission measurements are performed with nearly plane waves. To produce and detect them a two-dimensional horn antenna $a$ and a wave-guide antenna $b$ are placed, respectively, on the opposite sides of the crystal $240$~mm apart from it (see Fig.~\ref{fig3}). Both are screwed tightly to the top and bottom plate and are attached to waveguide-to-coaxial cable adapters. Transmission (reflection) spectra are measured with an Agilent PNA-L 5230 vectorial network analyzer (VNA). Microwave power is emitted into the region of the photonic crystal via one antenna $a$ and the magnitude and phase of the transmitted (reflected) power at the other (same) antenna $b$ ($a$) is determined relative to the input signal, thus yielding the complex valued scattering matrix element $S_{ba}(f)$. Systematic errors due to the attenuation of microwave power in the coaxial cables and the reflection at the connectors were removed by a proper calibration of the VNA. Ribbons of microwave absorption foam were placed between the plates at their edges to avoid disturbing reflections.
\begin{figure}[b]
{\includegraphics[width=\linewidth]{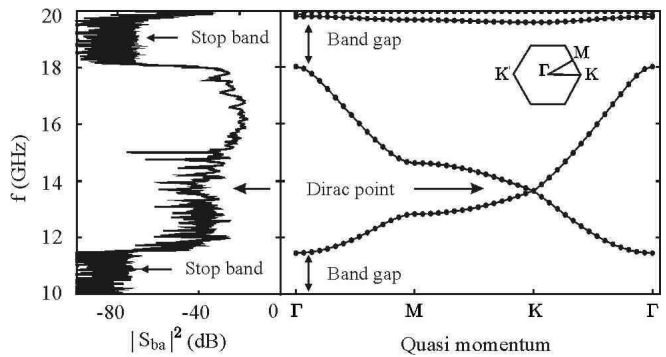}}
\caption{Left: The measured transmission spectrum from the horn to the wave guide antenna. Right: Numerically calulated band structure $f(\lvert \vec{k}\rvert)$ with  $\vec k$ chosen along the $\Gamma{\rm M}{\rm K}\Gamma$ path. The inset shows the Brillouin zone of the lattice and the irreducible Brillouin zone $\Gamma{\rm M}{\rm K}$.} 
\label{fig4}
\end{figure} 
\par The photonic crystal is arranged such, that the waves produced by the horn antenna impinge it in $\Gamma{\rm M}$ direction (see Figs.~\ref{fig1} and \ref{fig3}). The left part of Fig.~\ref{fig4} shows a measured transmission spectrum $|S_{ba}(f)|^2$  through the photonic crystal in $\Gamma{\rm M}$ direction. It shows two stop bands, one below 11.5~GHz and one between 18 and 19.8~GHz, where the transmission is reduced by 4 orders of magnitude. In between we observe transmission with a broad minimum around 14~GHz. The right part shows the calculated band structure along the path $\Gamma{\rm M}{\rm K}\Gamma$ (see inset of Fig.~\ref{fig4}) inside the first Brillouin zone. The first and second band touch at the corner ${\rm K}$ of the Brillouin zone. The positions of the experimental stop bands are in good agreement with the band gaps in the calculated band structure. The minimum in the transmission spectrum is located around the Dirac frequency and indicates a partial band gap \cite{Joannopoulos2008} between the first and second band expected for the transmission of plane waves in $\Gamma{\rm M}$ direction. However, since the incident wave is not perfectly plane and thus contains components with propagation direction different from the $\Gamma{\rm M}$ one, this band gap is not as well pronounced as the other two. Altogether the experimental transmission spectrum resembles the predicted band structure of the photonic crystal.

\par 

A very effective method to localize a Dirac point is to measure the frequency dependence of the local density of states (LDOS). It is defined as the imaginary part of the Green function for the photonic crystal, $L(\vec r,f)=-1/\pi{\rm Im}\;G(\vec{r},\vec{r},f)$. According to \cite{McPhedran2004} it can be written 
in terms of the frequencies $f_n(\vec{k})$ and the wave functions $\psi_n(\vec{k},\vec{r})$ of the $n$th propagating mode 
as  
\begin{equation}
 L(\vec{r},f)=\frac{1}{\mathcal{A}_{BZ}}\int_{BZ}\sum_{n}\vert\psi_n(\vec{k},\vec{r})\vert^2\frac{1}{2\pi}\delta(f-f_n(\vec{k}))d^2 k,
\label{Ldos}
\end{equation}
where the integration over the quasi momentum $\vec{k}$ extends over the first Brillouin zone with area $\mathcal{A}_{BZ}$.
Using the dispersion relation Eq.~(\ref{DiracDispersion}) and that the modulus squared of the wave function 
depends only weakly on $\vec{k}$ close to the Dirac point yields 
\begin{equation}
L(\vec{r}_a,f)\approx \frac{4\pi^2}{\mathcal{A}_{BZ}}\frac{\langle\vert \psi(\vec{r_a})\vert^2\rangle}{v_D}
\left\vert\frac{ f - f_D}{v_D}\right\vert\, . 
\label{LdosDirac}
\end{equation}
Here, $\langle\vert \psi(\vec{r_a})\vert^2\rangle$ denotes the average over $\vec{k}$ of the modulus squared of the wave functions in the vicinity of the Dirac point. Equation~(\ref{LdosDirac}) reflects the well-known theoretical result \cite{Wallace1947,Slonczewski1958,Castro2009} that near the Dirac frequency the LDOS tends to zero linearly in $f$. The LDOS can be determined  from the measured  reflection spectra $|S_{aa}(f)|^2$ with the antenna placed at a position $\vec{r}_a$ within the crystal. The connection between the LDOS and the measured reflection spectra is established via the scattering matrix formalism for microwave resonators \cite{Albeverio1996,Dietz2010} originally developed in the context of compound-nucleus reaction theory \cite{Mahaux1969}. Within this formalism the scattering matrix can be written in the form $\hat S(f)=(\openone -i\hat K)(\openone +i\hat K)^{-1}$, 
where $\hat K=\hat W^T(E-\hat H)^{-1}\hat W$ with  $\hat H$ denoting the Hamiltonian for the photonic crystal and $\hat W$ the matrix coupling the open channels, as e.g., the antenna channel to the modes of the photonic crystal. Expressing $\hat K$ in terms of the Green function of the photonic crystal, $G(\vec r_a,\vec r_b,E+i\epsilon)=\sum_n\psi^*_n(\vec r_b)\psi_n(\vec r_a)(E+i\epsilon-E_n)^{-1}$ leads to the relation

\begin{equation}
 1-\vert S_{aa}(f)\vert^2\propto\frac{L(\vec{r}_a,f)}{\vert\tilde{G}(\vec{r}_a,\vec{r}_a,f)+A\vert^2}.
\label{GreenSaa}
\end{equation}
Here, $\tilde{G}(\vec{r}_a,\vec{r}_a,f)$ denotes the renormalized Green function \cite{Tudorovskiy2008} and 
the constant $A$ depends on the coupling of the antenna to the photonic crystal.
In the vicinity of the Dirac point one can neglect the frequency dependence of the denominator, i.e., replace it by a constant $B(\vec{r}_a,f_D)$ and finally gets the relation between the experimental observable $\vert S_{aa}(f)\vert^2$ and the LDOS
\begin{equation}
 \vert S_{aa}(f)\vert^2\propto1-B(\vec{r}_a,f_D)L(\vec{r}_a,f)\, .
\label{ScatLdos}
\end{equation}
\begin{figure}[t]
{\includegraphics[width=\linewidth]{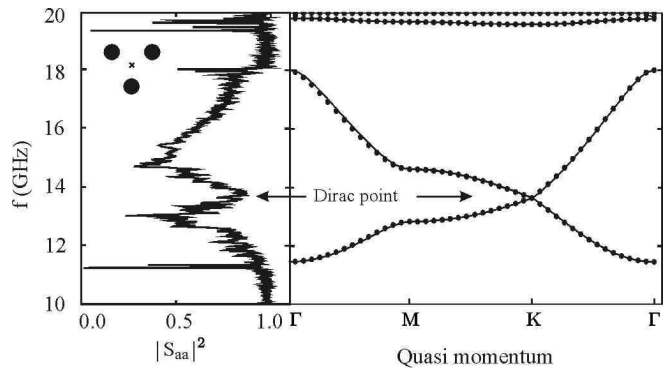}}
\caption{Left: A reflection spectrum measured with the wire antenna located in the middle of the crystal in the center of a triangle of cylinders as schematically shown in the inset. The sharp resonances at the edges of the bands are related to the so-called van Hove singularities. Right: For camparison the computed band diagram from Fig.~\ref{fig4} is shown.} 
\label{fig5}
\end{figure} 
Thus, according to Eq.~(\ref{LdosDirac}), the reflected power $\vert S_{aa}(f)\vert^2$ tends linearly to unity in the vicinity of the Dirac frequency. This is the ``fingerprint`` used to identify the Dirac point. 

\par For the measurement of the reflection spectra a wire antenna consisting of a metallic pin of 1~mm in diameter was lead through a 3~mm wide drilling in the top plate. The left part of Fig.~\ref{fig5} shows the reflection spectrum measured with an antenna placed in the middle of the photonic crystal as shown in the inset of this figure. The location of the antenna is chosen in the center of three cylinders forming a triangle to minimize the disturbance of the propagating mode at the Dirac frequency. The experimental reflection spectrum has a clearly pronounced maximum around $\approx$14~GHz, i.e., within the frequency range where the Dirac point is expected, and shows the characteristic cusp structure. The sharp resonances at the edges of the bands are related to the so-called van Hove singularities \cite{van1953}.

\par
\begin{figure}[t]
{\includegraphics[width=\linewidth]{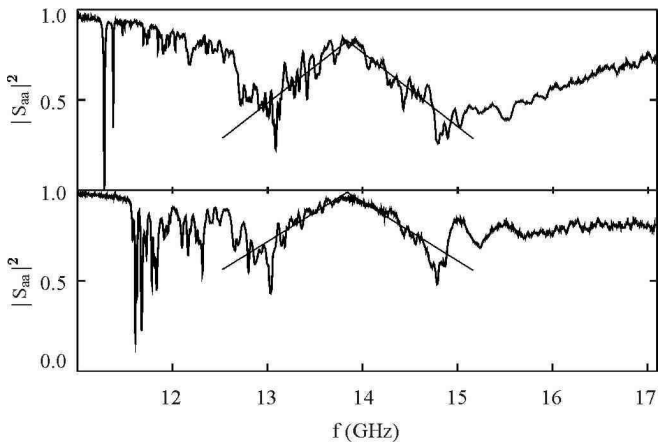}}
\caption{ Two measured reflection spectra in the vicinity of the Dirac frequency. The upper spectrum was measured with an antenna placed in the middle of the crystal, the lower one with an antenna placed 6 rows apart from the boundary of the crystal. The smooth solid lines are fits of Eq.~(\ref{fit}) to the experimental spectra.} 
\label{fig6}
\end{figure}

 The description of the experimental reflection spectrum around the Dirac frequency based on Eqs.~(\ref{LdosDirac}) and (\ref{ScatLdos}) allows to express the measured quantity $|S_{aa}(f)|^2$ in terms of a 3-parameter formula,
\begin{equation}
\vert S_{aa}(f)\vert^2=D-C\,\vert f-f_D\vert,
\label{fit}
\end{equation}
where the parameter $D$ takes into account that due to Ohmic losses the reflection deviates from unity even in the case of vanishingly small LDOS. The parameter $C$ describes the slope of the cusp in the reflection spectrum close to the Dirac frequency. The upper and the lower parts of Fig.~\ref{fig6} show the experimental reflection spectra in the frequency range 11-17~GHz for two different antennae together with the fit of Eq.~(\ref{fit}) to the data.
The spectrum shown in the upper panel of Fig.~\ref{fig6} was measured with an antenna placed in the middle of the crystal, that shown in the lower panel was obtained with one placed 6 rows apart from the crystal boundary. In the former case the antenna reached 2~mm into the space between the plates, in the latter only 0.5~mm. The upper spectrum is well described by Eq.~(\ref{fit}) and the values of the fit parameters are $D=0.826\pm0.003$, $C=0.413\pm0.006$, and the Dirac frequency $f_D=13.797\pm0.004$~GHz is in good agreement with the calculated value $f_D=13.81$~GHz. The group velocity $v_D$ at the Dirac point can not be extracted from $C$ because the value of $\langle\vert \psi(\vec{r_a})\vert^2\rangle$ in Eq.~(\ref{LdosDirac}) is unknown. The shape and position of the cusp are recovered in the lower reflection spectrum in Fig.~\ref{fig6}. The obtained values of the parameters are $D=0.988\pm0.001$, $C=0.329\pm0.003$ and $f_D=13.793\pm0.002$~GHz. The value for the Dirac frequency differs slightly from that determined for the antenna placed in the middle of the crystal but the difference is within the standard error of the fit. The value of the parameter $D$ is closer to unity in the latter case because the antenna reaches less into the space between the plates and thus couples weaker to the propagating modes. This is also reflected in the slope $C$ of the cusp, which is not as steep. Moreover, the cusp at the Dirac frequency is rounded in the lower spectrum. This is attributed to the proximity of the antenna to the crystal boundary. The fluctuations observed in the reflection spectra around the mean behaviour given by Eq.~(\ref{fit}) are attributed to quasi bound modes trapped in the photonic crystal. In the vicinity of the Dirac point, the measured reflection spectra closely resemble those obtained by conductance measurements in scanning tunneling microscopy of graphene flakes deposited on a graphite substrate \cite{Li2009}.

In summary, we have measured the transmission and the reflection spectra of a photonic crystal with a triangular lattice composed of metallic cylinders. A cusp structure was observed in the reflection spectra close to the expected Dirac frequency  and related to the local density of states in the photonic crystal, providing clear evidence for the existence of a Dirac point. The experimental setup presented in this work is well suited for the study of effects connected with so-called edge states in graphene \cite{Raghu2008,Ochiai2009} and the predicted pseudo-diffusive transmission at the Dirac point. However, the most direct application is the experimental investigation of properties of the eigenvalues and eigenfunctions of a Dirac billiard, by inserting the photonic crystal setup into a closed resonator box \cite{Wurm2009}.
\par The authors thank E. Bogomolny and T. Tudorovskiy for stimulating discussions, and the LPTMS in Orsay for its hospitality. This work has been supported by the DFG within the SFB 634.

\end{document}